\documentclass{elsart}
\usepackage{natbib}
\usepackage{graphicx}

\def\lsim{\lower.5ex\hbox{$\; \buildrel < \over \sim \;$}}
\def\gsim{\lower.5ex\hbox{$\; \buildrel > \over \sim \;$}}

\begin{document}
\runauthor{Georganopoulos \& Kazanas}
\begin{frontmatter}
\title{Decelerating  Flows in TeV Blazars}
\author{Markos Georganopoulos \& Demosthenes Kazanas}

\address{Laboratory for High Energy Astrophysics, NASA Goddard Space 
Flight Center, Code 661, Greenbelt, MD 20771, US}

\begin{abstract}

The peak of the de-absorbed energy distribution of the  TeV emitting 
blazars, all of the BL Lacertae (BL) class, can reach values up to 
$\sim 10$ TeV. In the context of synchrotron-self Compton (SSC) models
of  relativistic uniformly moving blobs of plasma, such high energy  
peak emission  can be reproduced only by assuming Doppler factors  of 
$\delta \sim 50$. However, such high values strongly disagree with the 
unification of FR I radio galaxies and BLs. Additionally, the recent 
detections of slow, possibly sub-luminal velocities in the sub-pc scale 
jets of the TeV BLs  MKN 421 and MKN 501 suggest that the jets in these 
sources decelerate  very early to mildly relativistic velocities 
($\Gamma\sim$ a few). In this work we examine the possibility that the 
relativistic flows in the TeV BLs are longitudinally decelerating. In this 
case, modest Lorentz factors ($\Gamma \sim 15$), decelerating down to  
values compatible with the recent radio interferometric observations, can 
reproduce the $\sim $ few TeV peak energy of these sources. Furthermore, 
such  decelerating flows are  shown to  reproduce the observed broadband 
BL - FR I luminosity ratios.

\end{abstract}

\end{frontmatter}

\section{Introduction}

The blazar TeV emission reaches the Earth partially  absorbed by the 
Diffuse InfraRed Background (DIRB) radiation which pair-produces with the 
TeV photons \cite{stecker92}. The de-absorbed  spectra depend on the 
source redshifts and the still elusive energy distribution of the DIRB.
However, for all expected DIRB forms, both the intrinsic peak energy 
$E_p$ and peak luminosity $L_p$ of the TeV spectral component are higher 
than those observed. Even for the nearby ($z=0.031$) MKN 421, $E_p$ can 
increase by a factor of $\sim 10$ after de-absorption to $\sim $ a few 
TeV \citep{dejager02}. The de-absorbed spectrum of   H1426+428 at z=0.129 
is even more extreme, characterized  by $E_p \sim 10$ TeV \cite{aharonian02}.

The BL synchrotron spectra exhibit a break at energies $\epsilon'_{b} 
\sim 10^{-4}-10^{-6}$ (primes denote energies on the flow rest frame, 
all energies normalized to $m_ec^2$) such that the greatest fraction of 
the synchrotron luminosity is at energies $\epsilon' > \epsilon'_{b}$. 
For this reason and because of the K-N decrease in the Compton scattering
cross section, electrons with $\gamma \gsim 1/\epsilon'_{b}$ will inverse 
Compton scatter a decreasing fraction of the available photons; as a result, 
the maximum IC luminosity will occur at energies $\epsilon'_p \simeq 1/
\epsilon'_{b}$, or $\epsilon'_{p}\epsilon'_{b}\simeq 1$. The values of 
$\epsilon'_p, \, \epsilon'_{b}$ observed at the lab frame are $\epsilon_p 
= \delta \epsilon'_p$ and $\epsilon_{b} = \delta \epsilon'_{b}$, yielding 
the following relation between $\delta$ and the observed energies 
$\epsilon_p, \, \epsilon_{b}$ 
\begin{equation}
\delta \gsim (\epsilon_{b}\; \epsilon_{p})^{1/2} =   40\;(\nu_{b,16} 
\;E_{p,\,10\,\rm TeV})^{1/2},  \label{d_constr}
\end{equation} 
where $\nu_{b,16}$ is the observed synchrotron break frequency in units
of $10^{16}$ Hz, $E_{p,\,10\,\rm TeV}$ is the energy of the de-absorbed
TeV  peak  in units of 10 TeV and $\delta=1/\Gamma(1-\beta\cos\theta)$,  
with $\beta$ the dimensionless flow speed and $\theta$ its 
angle to the observer's line of sight. These values of $\delta (\simeq 
\Gamma)$  are in strong conflict \citep{chiaberge00} with the unification 
of BLs and FR I radio galaxies \citep{urry95}, which require Lorentz factors 
$\Gamma \sim 3-7$. Additional constrains for the inner jet flow of the TeV 
blazars come from the small, possibly sub-luminal apparent  velocities 
observed interferometrically  in the sub-pc scale jet of  MKN 421 and 501 
\citep{marscher99,piner99,edwards02}, suggestive of a  decelerating 
 flow  in the inner jet of TeV blazars \citep{marscher99}.

\section{Decelerating relativistic flows:  application in TeV blazars}

Motivated by the above analysis and the unification of the multiwavelength
spectra of the hotspots of FR II radio galaxies and quasars in terms of
decelerating relativistic flows \cite{georganopoulos03}, we propose that 
the same considerations are applicable in resolving the conflict of the high 
$\delta$'s with TeV blazar unification. So, we assume that a power law electron
distribution is injected at the base of a relativistic flow that decelerates
while at the same time cooling radiatively. The highest frequencies of its  
synchrotron component originate preferentially at its fast base where the 
electrons are more energetic and its Lorentz factor largest. As both 
the flow velocity and electron  energy drop with radius, the locally emitted 
synchrotron spectrum shifts to lower energies while its beaming pattern 
becomes wider. The observed synchrotron spectrum is the convolution of the 
comoving emission from each radius weighted by the beaming amplification 
at each radius. At small angles the observed spectrum is harder than that
observed at larger angles. This is the result of the progressively smaller 
$\Gamma$ of the flow with distance in combination with the concomitant 
lower electron energies due to cooling.


The inverse Compton emission of such a flow behaves in a more involved
way: Electrons will upscatter the locally produced synchrotron seed photons,
giving rise to a local SSC emission with $\delta-$dependence similar to 
to that of synchrotron. However, the electrons of a given radius scatter 
not only the locally produced synchrotron photons, but also those produced 
downstream in the flow. The energy density of the latter, will appear 
Doppler boosted in the fast (upstream) part of the flow by $\sim 
\Gamma_{rel}^2$ \citep{dermer95}, where $\Gamma_{rel}$ is the relative 
Lorentz factor between the fast and slow part of the flow. With their
maximum energy being lower (because of cooling) and their energy density
amplified they can now contribute to the IC emission at energies higher 
than expected on the basis of uniform velocity models (see section 1). 
Also the $\delta-$dependence of this {\sl upstream Compton (UC)} emission 
will be different from that of SSC and more akin to that of external
Compton (EC) \cite{dermer95}.

\begin{figure*}

\centerline{
\includegraphics[width=2.75in]{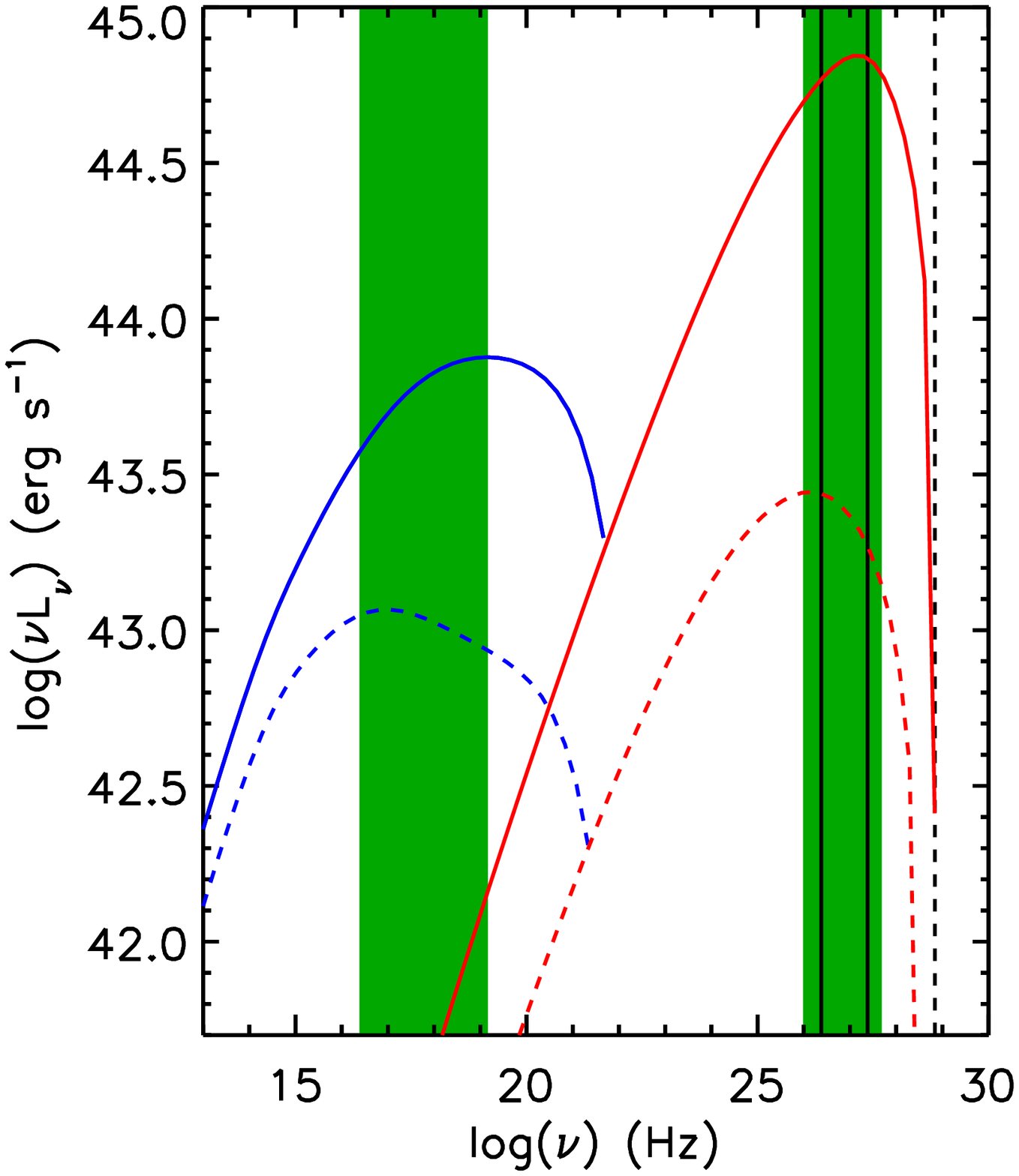}
\includegraphics[width=2.75in]{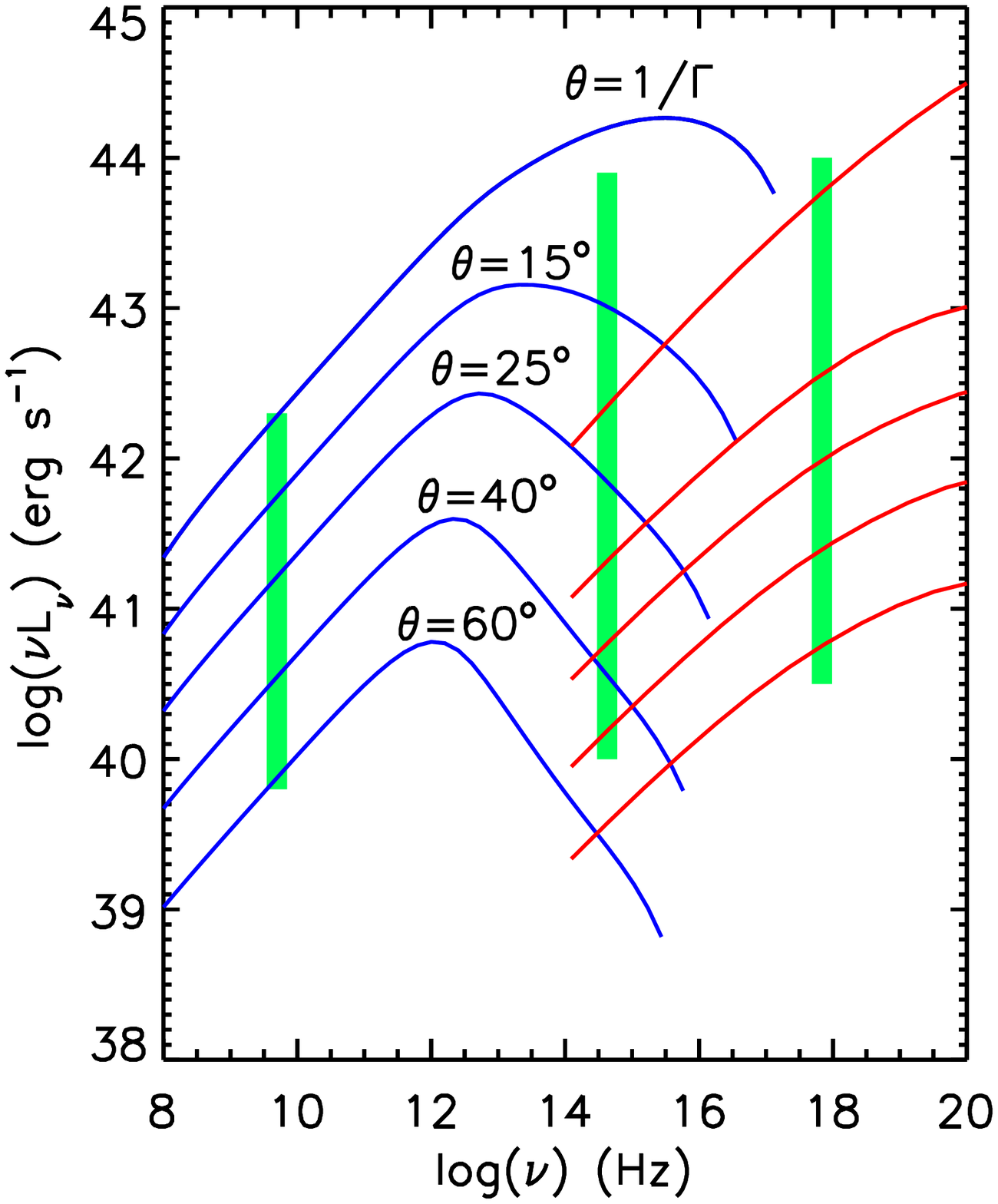}
}
\caption{{\sl Left panel:} the Synchrotron and IC emission 
from a decelerating  relativistic flow $\Gamma=15$ to $\Gamma=4$
at $\theta=3^{\circ}$ (solid line) and $\theta=6^{\circ}$ (broken 
line) (see text for details). The broken vertical line corresponds 
to the maximum IC energy $\gamma_{max}\,\delta$ for 
$\theta=3^{\circ}$. The shaded areas correspond approximately to the energy 
range of X-ray and TeV telescopes and the two vertical solid lines energies
to  1 and 10 TeV. {\sl Right panel:} The SED of a decelerating flow for a 
range of observing angles.  The physical parameters are the same as before, 
but with $\gamma_{max}=2\times 10^{5}$. The shaded bars reflect the 
average luminosity range in radio, optical and X-rays, between the 
samples of BLs and FR I radio galaxies \cite{trussoni03}.}\label{f1}
\end{figure*} 

In the left panel of fig. 1 we plot the SED for a flow decelerating from 
$\Gamma=15$ to $\Gamma=4$ for two angles $\theta=3^{\circ}$ and $\theta=
6^{\circ}$ over a distance $Z = 2\times 10^{16}$ cm. The radius of the cylindrical
flow is set to $R = Z = 2\times 10^{16}$ cm, while the electron distribution
at the base of the flow is $n(\gamma)\propto \gamma^{-2}$, $\gamma \leq 3\times 
10^7$ and the magnetic field $B=0.1$ G,  half its equipartition value. 
At $\theta=3^{\circ}$ this model achieves a peak energy for the 
high energy  component at $\sim 10$ TeV, using a modest initial Lorentz factor 
of $\Gamma=15$. Note the strong dependence of synchrotron peak energy 
on $\theta$ discussed above. Finally, note that, in contrast to uniform 
velocity homogeneous SSC models, the Compton component is more sensitive 
to orientation than synchrotron, an indication that UC scattering dominates 
the $\sim $ TeV observed luminosity.

We now turn our attention to the problem of the unification of BLs with FR I
sources. A comparison of a sample of FR I nuclei to BLs of similar extended 
radio power shows that  FR I's  are overluminous by a factor of $10-10^4$
compared to their expected luminosity \cite{chiaberge00,trussoni03}, under 
the assumption that BLs are  characterized by Lorentz factors $\Gamma\sim15$, 
and that FR I's are seen under an average angle of $60^{\circ}$. In particular 
the average BL to FR I nucleus luminosity ratio at radio, optical and X-ray 
bands was found to be: $\log (L_{BL}/L_{FR\;I})_R \approx 2.4$,  $\log (L_{BL}
/L_{FR\;I})_{opt} \approx 3.9$,  $\log (L_{BL}/L_{FR\;I})_{X} \approx 3.5$.
In the right panel of fig. 1 we plot as vertical bars the luminosity 
separation of BLs and FR Is according to \cite{chiaberge00, trussoni03}.
We also plot the SED of a decelerating flow with physical parameters similar 
to the one described above but with $\gamma_{max}= 2\times 10^{5}$ to
produce an SED synchrotron peak similar to those of the intermediate BLs 
that correspond in extended radio power to the FR Is of \cite{chiaberge00,
trussoni03}. As can be seen, the range of the model SEDs between  
$\theta=60^{\circ}$ (FR I) and  $\theta=1/\Gamma$ (BL) reproduce 
well the observed range in luminosities, while a uniform velocity model of 
$\Gamma = 15$ (as demanded by fits of the $\gamma-$ray spectra) would 
produce at $\theta=60^{\circ}$ an SED many orders of magnitude smaller
than shown in the figure.

\appendix
\end{document}